\documentclass[epj]{svjour}
\usepackage{amssymb}
\usepackage{amsmath}
\usepackage{graphics}
\begin{document}
\title{State transfer in dissipative and dephasing environments}

\author{M.L. Hu\thanks
{e-mail: mingliang0301@xupt.edu.cn}
}
\institute{School of Science, Xi'an University of Posts and
           Telecommunications, Xi'an 710061, China}
\date{Received: date / Revised version: date}
%
\abstract{ By diagonalization of a generalized superoperator for
solving the master equation, we investigated effects of dissipative
and dephasing environments on quantum state transfer, as well as
entanglement distribution and creation in spin networks. Our results
revealed that under the condition of the same decoherence rate
$\gamma$, the detrimental effects of the dissipative environment are
more severe than that of the dephasing environment. Beside this, the
critical time $t_c$ at which the transfer fidelity and the
concurrence attain their maxima arrives at the asymptotic value
$t_0=\pi/2\lambda$ quickly as the spin chain length $N$ increases.
The transfer fidelity of an excitation at time $t_0$ is independent
of $N$ when the system subjects to dissipative environment, while it
decreases as $N$ increases when the system subjects to dephasing
environment. The average fidelity displays three different patterns
corresponding to $N=4r+1$, $N=4r-1$ and $N=2r$. For each pattern,
the average fidelity at time $t_0$ is independent of $r$ when the
system subjects to dissipative environment, and decreases as $r$
increases when the system subjects to dephasing environment. The
maximum concurrence also decreases as $N$ increases, and when
$N\rightarrow\infty$, it arrives at an asymptotic value determined
by the decoherence rate $\gamma$ and the structure of the spin
network.
} 
\maketitle
\section{Introduction}
\label{intro} An important and new emerging pursuit of quantum
information processing (QIP) task is the high-fidelity transmission
of quantum states across a pre-engineered quantum spin network. This
is so because when performing any QIP tasks, one needs to exchange
quantum information between distant nodes (e.g., the core processor,
storage, etc.) of a quantum computer. In general, the basic idea of
transmitting a quantum state from one location of a network to
another proceeds in three steps \cite{Ref1}. The first step is the
initialization of the spin medium in a fiducial pure state, then the
sender Alice encodes the state needs to be transmitted at one node
of the network, and time evolution of the system for a proper
interval of time, finally, the state will be recovered at the nodes
belonging to the receiver Bob with certain fidelity. Particularly,
the quantum channel may enable state transfer with fidelity better
than any classical communication channel.\footnote{The context of a
footnote.}

Since the seminal work of Bose \cite{Ref1}, in which the author
demonstrated that an unmodulated ferromagnetic spin chain with
Heisenberg interactions can be used as a channel for short distance
quantum communication, there has been an increasing discussion of
quantum state transfer in various physical systems in the
literature. Among such systems, the solid-state quantum system with
spin-spin interactions offers an excellent theoretical framework for
designing networks for perfect state transfer (PST)
\cite{Ref1,Ref2,Ref3,Ref4,Ref5,Ref6,Ref7,Ref8,Ref9,Ref10,Ref11,Ref12,Ref13,Ref14,Ref15,Ref16}.
Particularly, Christandl et al. suggested a PST algorithm which can
transfer an arbitrary quantum state between the opposite ends of a
spin chain or the two antipodes of the one-link and the two-link
hypercubes \cite{Ref3,Ref4}. Zhang and Long et al. \cite{Ref13}
realized this PST algorithm in a three-qubit XX spin chain using
liquid nuclear magnetic resonance (NMR) system. Since then, quantum
spin networks as an ideal communication channel to realize QIP tasks
have been extensively studied. Shi et al. \cite{Ref6} presented a
series of perfect spin channels according to the
spectrum-parity-matching condition they derived. Then, Kostak et al.
\cite{Ref14} established a general formalism for engineering spin
Hamiltonians for PST in spin networks of arbitrary topology and
coupling configuration. Christandl's innovative works
\cite{Ref3,Ref4} were extended by Jafarizadeh and Sufiani in a
recent work \cite{Ref15}, in which they adopted distance-regular
graphs as spin networks and found that any such network can achieve
unit fidelity of state transfer over arbitrarily long distances.
Moreover, D'Amico et al. \cite{Ref10} showed that one can create and
distribute entanglement with an interaction-modulated Y-shaped spin
network, particularly, with a slightly complicated bifurcation
structure, one can even froze the entanglement by applying a phase
flip to one spin out of each pair.

Apart from the aforementioned protocols which mainly concentrated on
quantum spin chains with only nearest-neighbor (NN) couplings, in
Ref. \cite{Ref17} Paternostro et al. presented another scheme which
enables nearly optimal state transfer in imperfect artificial spin
networks with all the qubits are mutually coupled. Then Kay
demonstrated that PST is also possible in the presence of
next-nearest-neighbor (NNN) couplings \cite{Ref18}. Moreover,
compared to the case where the system contains only two-spin
interactions, the authors in Ref. \cite{Ref19} showed that the speed
of quantum state transfer across the XY spin chain can be
significantly increased by introducing the three-spin interaction.
Besides the spin-1/2 systems, state transfer efficiency of a
bilinear-biquadratic (BB) spin-1 Heisenberg chain has also been
discussed \cite{Ref20}.

Most recently, Franco et al. \cite{Ref21} presented a
control-limited scheme \cite{Ref22} for PST through a pre-engineered
spin chain. Different from the previous schemes whose achievements
relies crucially on the initialization of the channel state, Franco
et al. \cite{Ref21} demonstrated that such state initialization is
in fact inessential to the performance of the protocol if proper
encoding at the end of the chain is performed. The key requirements
for their scheme are the arrangement of proper time evolution and
the performance of clean projective measurements on the two end
spins, which considerably relaxes the prerequisites for obtaining
reliable state transfer across interacting-spin systems. Motivated
by this innovative work, Markiewicz and Wie\'{s}niak \cite{Ref23}
proposed a special type of encoding strategy for PST of a two-qubit
state, where no remote-cooperated global state initialization and
any additional communication are needed.

Besides these exciting progresses in this direction, however, one
cannot neglect the fact that real quantum system is very susceptible
to decoherence
\cite{Ref24,Ref25,Ref26,Ref27,Ref28,Ref29,Ref30,Ref31,Ref32,Ref33,Ref34,Ref35}.
For nearly all QIP tasks, the quantum system of interest will
inevitably be interacted with its surrounding environments (or the
thermal reservoirs). This unavoidable mutual interaction often cause
the initial state of the system becomes entangled with the
environments in an uncontrollable way, and it is this entanglement
of the system with the environment that induces decoherence. The
decoherence can affects quantum interferences of quantum systems and
leads to the degradation of quantum coherence, and thus becomes a
serious limiting factor baffling the physical realization of a
quantum computer. Since the system of interest is inevitably subject
to decoherence and decay processes, no matter how much they may be
screened from the external environments, it would seem important to
consider possible methods to minimize, delay, or even eliminate this
unwanted detrimental effects of environments in the practical
realization of various QIP tasks \cite{Ref36,Ref37}.

Dissipation and dephasing are two important sources of decoherence
in practice, and their effects on entanglement dynamics have been
analyzed very recently in Refs. \cite{Ref26,Ref27,Ref28,Ref29} for
the special case of two spin-halves where an analytical treatment is
possible. The results revealed that the entanglement dynamics
depends not only on the parameters of the system, but also on the
system-environment coupling strength and the initial states of
concern. How these decoherence modes affect quantum state transfer
fidelity in a spin chain is an interesting and urgent problem needs
to be solved, and it is also of both theoretical and experimental
significance. We will address this question in the present paper.
The contents are organized as follows. In Section 2, we introduced
the technique for solving the Lindblad form of the master equation.
Based on this technique, we obtained the time evolution operator of
the system and the explicit form of the state transfer fidelity.
Then in Section 3, we investigated dynamics of the state transfer
fidelity with the system subjects to dissipative and dephasing
environments, where we concentrated on several ideal spin channels
in the absence of decoherence, and aimed at revealing the extent to
which the decoherence affects quantum state transfer. Section 4 is
devoted to the analysis of the effects of the dissipative and
dephasing environments on entanglement distribution and creation
between distant parties through a pre-engineered spin network.
Finally, we concluded this paper with a short discussion in Section
5.

\section{Solution of the master equation}
\label{sec:2} As pointed in Section 1, the unavoidable mutual
interaction of an open quantum system with its surrounding
environment is an important source of decoherence. In order to
describe such phenomena, a master equation approach can be used. In
the present work, we investigate how the dissipative and the
dephasing environments affect quantum state transfer, as well as
entanglement distribution and creation through a pre-engineered spin
network. We assume that during the decoherence processes each spin
of the system interacts only, and independently, with its own
environment (this assumption is legitimate provided the constituents
composing the quantum system are separated by distances large
enough, however, it may not be totally valid in the case where the
spins are close enough to each other to couple strongly enough),
then in the Markovian limit, the dynamics of the system can be
described by a general quantum master equation (ME) of the Lindblad
form \cite{Ref38,Ref39}
\begin{equation}
 {d\rho\over dt}=-i[\hat{H},\rho]+\sum_n\mathcal{L}_n\rho,
\end{equation}
where $\hat{H}$ and $\rho$ denote, respectively, the Hamiltonian and
the density operator of the system, and the summation on the second
term of the right-hand side of this equation runs over all the spins
involved. $\mathcal {L}_n$ refers to the Lindblad superoperator,
which describes the independent interaction between each spin and
the environment and can be written, under weak system-environment
interaction and in the Markovian limit, as \cite{Ref38}
\begin{equation}
 \mathcal{L}_n\rho={\gamma\over2}(2c_n\rho c_n^\dag-c_n^\dag
 c_n\rho-\rho c_n^\dag c_n).
\end{equation}
Here $\gamma$ is the phenomenological parameter that describes the
coupling strength of a qubit with its local environment. The
system-environment coupling operator is given by $c_n=\sigma_n^-$
for the dissipative environment, and $c_n=\sigma_n^+\sigma_n^-$ for
the dephasing environment. The first term on the right-hand side of
Eq. (1) generates a coherent unitary evolution of the system, while
the second term represents the decoherence effects of the
environment on the system and generates an incoherent dynamics of
the system.

In order to assess the extent to which the decoherence affects
dynamics of a correlated spin system, one must first try to obtain
solutions of the ME. Although for systems with certain symmetries,
the ME can be solved by adopting the superoperator method
\cite{Ref40} or the Lie algebraic method (see, e.g., Ref.
\cite{Ref41} and references therein). For general cases, however,
this is still an uneasy task because the ME expressed in Eq. (1)
contains terms coupled with each other in a non-trivial way.
Although one can convert it to an equivalent $c$-number partial
differential equation \cite{Ref42} and then solve it numerically
using the conventional Runge-kutta algorithm; however, for system
with large number of spins, this is still a troublesome task for it
wasting too much computation resources to obtain the density
operator $\rho(t)$, and the situation becomes even more cumbersome
when one performs simulations of $\rho(t)$ with different initial
states because for every initial state one must perform the
Runge-kutta algorithm again to obtain the solutions of the ME. In
order to circumvent this quandary, in the present paper we resort to
another technique to solve the ME, as one will see, the success of
this technique depends solely on the exact diagonalization of a
generalized superoperator $\tilde{\Lambda}$, and it has no relation
with the initial state of the system. This technique provides an
efficient way for the investigation of the dynamics of the system
and not only allow for an efficient numerical analysis, but also, in
some cases, for exact algebraic solutions.

The basic idea of this technique is as follows. Instead of solving
the ME directly, one can first define two ancillary operators
$\tilde{\Lambda}$ and $\tilde{\rho}$, which satisfy the following
relations
\begin{equation}
 {d\tilde{\rho}\over{dt}}=\tilde{\Lambda}\tilde{\rho},
\end{equation}
where
\begin{equation}
 \tilde{\rho}=(\rho_{1,:},\rho_{2,:},\ldots,\rho_{d,:})^T.
\end{equation}
Here $d$ denotes the dimension of the density matrix $\rho$,
$(\ldots)^T$ denotes the transpose of $(\ldots)$, and
$\rho_{n,:}=[\rho_{n1},\rho_{n2},\\\ldots,\rho_{nd}]$ denotes the
elements lie in the $n$th row of $\rho$. After defining these two
ancillary operators, the formal solution of Eq. (3) can be expressed
explicitly as
\begin{equation}
 \tilde{\rho}(t)=\tilde{U}(t)\tilde{\rho}(0).
\end{equation}
Since $\rho$ is a $d\times d$ density matrix, the time evolution
operator $\tilde{U}(t)=\exp(\tilde{\Lambda}t)$ is a $d^2\times d^2$
matrix, and the initial state of the system $\tilde{\rho}(0)$ is a
$d^2\times 1$ column vector. Moreover, from the definition of
$\tilde{\rho}(t)$ one can obtain the equality
$\tilde{\rho}_{(i-1)d+j}(t)=\rho_{i,j}(t)$.

Now the key problem is how to obtain the explicit expression of
$\tilde{\Lambda}$. The crucial observation to this problem is that
the right-hand side of the ME expressed in Eq. (1) can be divided
into the following three terms, i.e.
\begin{equation}
 -i[\hat{H},\rho]+\sum_n\mathcal{L}_n\rho=A\rho+\rho B+\gamma\sum_n
 c_n\rho c_n^\dag,
\end{equation}
where $A$ and $B$ are operators which can be obtained directly from
Eqs. (1) and (2) as $A=-i\hat{H}-\gamma\sum_n(c_n^\dag c_n/2)$,
$B=i\hat{H}-\gamma\sum_n(c_n^\dag c_n/2)$. Hereafter we adopt the
conventional notation $M_{ij}$ to denote the element lies in the
$i$th row and $j$th column of a matrix $M$, and $V_i$ denotes the
$i$th element of a column vector $V$. In order to express the ME in
forms of $d\tilde{\rho}/dt=\tilde{\Lambda}\tilde{\rho}$, we make the
following three transformations:
$A\rho\mapsto\tilde{\Lambda}^{(1)}\tilde{\rho}$, $\rho
B\mapsto\tilde{\Lambda}^{(2)}\tilde{\rho}$ and $\gamma
\sum_{n}c_n\rho c_n^\dag\mapsto\tilde{\Lambda}^{(3)}\tilde{\rho}$.
To ensure the achievement of the first transformation
$A\rho\mapsto\tilde{\Lambda}^{(1)}\tilde{\rho}$, the equality
$(A\rho)_{ij}=(\tilde{\Lambda}^{(1)}\tilde{\rho})_{(i-1)d+j}$ must
be hold, or equivalently,
$\sum_{k}A_{i,k}\rho_{k,j}=\tilde{\Lambda}_{(i-1)d+j,:}^{(1)}\tilde{\rho}$,
then after a straightforward algebra, one can obtain
$\tilde{\Lambda}^{(1)}=A\otimes I$. Similarly, for the second
transformation, one can obtain $\tilde{\Lambda}^{(2)}=I\otimes B'$,
while for the third transformation, one can obtain
$\tilde{\Lambda}^{(3)}=\gamma\sum_{n}c_{n}\otimes c_{n}^{*}$. In the
above expressions, $B'$ denotes the transposed matrix of $B$,
$c_{n}^{*}$ denotes the conjugate of $c_n$, and $I$ is the $d\times
d$ identity matrix. With the help of these results, Eq. (3) becomes
\begin{equation}
 {d\tilde{\rho}\over{dt}}=\sum_{n=1}^{3}\tilde{\Lambda}^{(n)}\tilde{\rho}.
\end{equation}

This is just the desired form of the transformed master equation,
where the newly defined operator is given by
$\tilde{\Lambda}=A\otimes I+I\otimes B'+\gamma\sum_{n}c_{n}\otimes
c_n^*$. Once the explicit form of $\tilde{\Lambda}$ has been found,
the time evolution operator $\tilde{U}(t)=\exp{(\tilde{\Lambda}t)}$
can be formed and the formal solution of the master equation is
obtained, which can be expressed as Eq. (5). To obtain the density
operator at arbitrary time $t$, one only needs to apply the time
evolution operator $\tilde{U}(t)$ to the initial state
$\tilde{\rho}(0)$ of the system.

Now we try to convert the matrix exponential form of the time
evolution operator $\tilde{U}(t)=\exp{(\tilde{\Lambda}t)}$ to the
matrix form. For this purpose, we expanded it in terms of the Taylor
series as
\begin{equation}
 \exp{(\tilde{\Lambda}t)}=\sum_{n}{(\tilde{\Lambda}t)^n\over n!},
\end{equation}
where $n!$ denotes the factorial of $n$. If the $d^2\times d^2$
matrix $\tilde{\Lambda}$ is diagonalizable ($\tilde{\Lambda}$ cannot
always be diagonalized since it is generally complex and
non-Hermitian), i.e., there exists an invertible matrix $M$ that
satisfying the relation $M^{-1}\tilde{\Lambda}M=diag(\mathcal{E}_1,
\mathcal{E}_2,\ldots,\mathcal{E}_{d^2})$, with the diagonal entries
$\mathcal{E}_k$ $(k=1,2,\ldots,d^2)$ being the eigenvalues of the
operator $\tilde{\Lambda}$, then from Eq. (8) one can obtain
\begin{equation}
 M^{-1}\exp{(\tilde{\Lambda}t)}M=\sum_n{(M^{-1}\tilde{\Lambda}tM)^n\over
 n!}=\exp{(\mathcal{E}t)}.
\end{equation}
From Eq. (9) one can obtain directly that the matrix form of the
time evolution operator $\tilde{U}(t)$ is given by
\begin{equation}
 \tilde{U}(t)=M\exp{(\mathcal{E}t)}M^{-1},
\end{equation}
where $M^{-1}$ represents the inverse matrix of $M$, and $M$ is the
matrix with the eigenvectors of $\tilde{\Lambda}$ as its columns.

From the above analysis, one can see that the achievements of this
technique depends solely on the exact diagonalization of the
operator $\tilde{\Lambda}$, once $\tilde{\Lambda}$ is diagonalized,
one can obtain $\tilde{\rho}(t)$ as well as the matrix form of
$\rho(t)$ at any time $t$ by applying the time evolution operator
$\tilde{U}(t)$, particularly, $\tilde{U}(t)$ has no relation with
the initial state $\tilde{\rho}(0)$ of the system, which is
economical for performing numerical calculations with different
initial states, and this is different from that of the Runge-kutta
algorithm.

In the present paper, we first consider effects of dissipative and
dephasing environments on state transfer in a spin chain. We assume
the quantum state to be transmitted is encoded at the $m$th spin as
$|\varphi_{in}\rangle=\cos{(\theta/2)}|0\rangle+e^{i\phi}\sin{(\theta/2)}|1\rangle$
(with $|0\rangle$ and $|1\rangle$ represent the state of spin up and
down, respectively), and all the other spins in the chain are
initialized to the ground state $|0\rangle$, thus the initial state
of the whole system at time $t=0$ becomes
\begin{equation}
 |\psi(0)\rangle=\cos{\theta\over 2}|\textbf{0}\rangle+e^{i\phi}\sin{\theta\over
 2}|m\rangle,
\end{equation}
where $|\textbf{0}\rangle=|00\ldots0\rangle$, $|m\rangle=\sigma_m^+
|0\rangle^{\otimes N}$, $0\leqslant\theta\leqslant\pi$ and
$0\leqslant\phi\leqslant 2\pi$ are the relative phase angles. For
this type of initial state, the nonzero elements of
$\tilde{\rho}(0)$ can be written as
\begin{eqnarray}
 &\tilde{\rho}_{Nm-N+2m-1}(0)=\sin^2{\theta\over 2},\quad
  \tilde{\rho}_{N^2+2N+1}(0)=\cos^2{\theta\over 2},\nonumber\\
 &\tilde{\rho}_{Nm+m}(0)={1\over 2}e^{i\phi}\sin\theta,\quad
  \tilde{\rho}_{N^2+N+m}(0)=\tilde{\rho}_{Nm+m}^\dag(0).\nonumber\\
\end{eqnarray}
Except these four elements, all the other elements of
$\tilde{\rho}(0)$ are zero. Here $N$ denotes the length of the
chain, and we have used the fact that for initial state
$|\psi(0)\rangle$ with the system subjects to dissipative and
dephasing environments, its dynamics is completely determined by the
time evolution in the zero and single excitation subspace
$\mathcal{H}_{0\oplus 1}$ \cite{Ref26,Ref27,Ref28,Ref29}, thus it
suffices to restrict our attention to the dynamics of $\rho(t)$ in
this $(N +1)$-dimensional subspace spanned by $\{|\textbf{0}\rangle,
|1\rangle, |2\rangle,\ldots,|N\rangle\}$ which greatly facilitates
the following computation process.

To evaluate the extent to which the decoherence affects state
transfer in a spin chain, we adopt the concept of fidelity
\cite{Ref1} $f=\langle \varphi_{in}|\rho_{n}(t)|\varphi_{in}\rangle$
as an estimation of the quality of the state transfer from the
sender, conventionally named Alice, to the receiver Bob, where
$\rho_{n}(t)$ is the single qubit reduced density matrix. The
fidelity measures the overlap between the input state and the output
state. In general, if one encodes the state needs to be transmitted
at one end of the chain, then after some time $t$, the state will be
recovered at another end of the chain automatically, with however,
the fidelity $f<1$. With elaborately designed structures of the spin
chain which is ideally protected from its surrounding environments,
the maximum fidelity may reach unity
\cite{Ref3,Ref4,Ref6,Ref14,Ref15}. When subjects to decoherence
environments, as can be seen in the following sections, the transfer
fidelity cannot reach unity even with elaborately designed spin
structures.

In the standard basis $\{|0\rangle, |1\rangle\}$, the single qubit
reduced density matrix $\rho_{n}(t)$ can be obtained by tracing out
all other qubits except $n$ from $\rho(t)$ as
\begin{equation}
 \rho_{n}(t)=\left(\begin{array}{cc}
  1-\rho_{nn}(t)  & \rho_{N+1,n}(t) \\
  \rho_{n,N+1}(t) & \rho_{nn}(t)
 \end{array}\right).
\end{equation}
Here $\rho_{mn}(t)$ denotes the element lies in the $m$th row and
$n$th column of $\rho(t)$. Then the state transfer fidelity from the
$m$th qubit to the $n$th qubit can be computed as
\begin{equation}
 f=\cos^{2}{\theta\over 2}-\rho_{nn}\cos\theta+{1\over
 2}(e^{i\phi}\rho_{N+1,n}+e^{-i\phi}\rho_{n,N+1})\sin\theta.
\end{equation}
In terms of $\tilde{\rho}(t)$ the elements of $\rho_{n}(t)$ can be
written as
\begin{eqnarray}
 &\rho_{nn}(t)=\tilde{\rho}_{Nn-N+2n-1}(t),\nonumber\\
 &\rho_{n,N+1}(t)=\tilde{\rho}_{Nn+n}(t),\nonumber\\
 &\rho_{N+1,n}(t)=\tilde{\rho}_{N^2+N+n}(t).
\end{eqnarray}

The explicit form of $\tilde{\rho}_{Nn-N+2n-1}(t)$,
$\tilde{\rho}_{Nn+n}(t)$ and $\tilde{\rho}_{N^2+N+n}(t)$ can be
expressed as
\begin{eqnarray}
 \tilde{\rho}_{Nn-N+2n-1}&=&\tilde{U}_{Nn-N+2n-1,Nm-N+2m-1}\sin^2{\theta\over 2}\nonumber\\&&
  +\tilde{U}_{Nn-N+2n-1,N^2+2N+1}\cos^2{\theta\over 2}\nonumber\\&&
  +{1\over 2}(\tilde{U}_{Nn-N+2n-1,Nm+m}e^{i\phi}\nonumber\\&&
  +\tilde{U}_{Nn-N+2n-1,N^2+N+m}e^{-i\phi})\sin\theta,\nonumber\\
 \tilde{\rho}_{Nn+n}&=&\tilde{U}_{Nn+n,Nm-N+2m-1}\sin^2{\theta\over 2}\nonumber\\&&
  +\tilde{U}_{Nn+n,N^2+2N+1}\cos^2{\theta\over 2}\nonumber\\&&
  +{1\over 2}(\tilde{U}_{Nn+n,Nm+m}e^{i\phi}\nonumber\\&&
  +\tilde{U}_{Nn+n,N^2+N+m}e^{-i\phi})\sin\theta,\nonumber\\
 \tilde{\rho}_{N^2+N+n}&=&\tilde{U}_{N^2+N+n,Nm-N+2m-1}\sin^2{\theta\over 2}\nonumber\\&&
  +\tilde{U}_{N^2+N+n,N^2+2N+1}\cos^2{\theta\over 2}\nonumber\\&&
  +{1\over 2}(\tilde{U}_{N^2+N+n,Nm+m}e^{i\phi}\nonumber\\&&
  +\tilde{U}_{N^2+N+n,N^2+N+m}e^{-i\phi})\sin\theta.\nonumber\\
\end{eqnarray}

Substituting these results into Eq. (14), one can obtain the
explicit form of the state transfer fidelity. Particularly, when
$\theta=\pi$, we obtain the transfer fidelity of an excitation as
$f=\langle 1|\rho_n(t)|1\rangle=\tilde{U}_{Nn-N+2n-1,Nm-N+2m-1}$. On
the other hand, in order to assess the efficiency of the quantum
spin channel of interest, it is more beneficial to calculate the
average fidelity (the fidelity averaged over all pure input states
in the Bloch sphere) $F={1\over 4\pi}\int fd\Omega={1\over
4\pi}\int_0^\pi\int_0^{2\pi}f\sin\theta d\theta d\phi$ \cite{Ref1},
which can be computed as
\begin{eqnarray}
 F&=&{\tilde{U}_{N^2+N+n,N^2+N+m}+\tilde{U}_{Nn+n,Nm+m}\over 6}\nonumber\\&&
     +{\tilde{U}_{Nn-N+2n-1,Nm-N+2m-1}\over 6}\nonumber\\&&
     -{\tilde{U}_{Nn-N+2n-1,N^2+2N+1}\over 6}+{1\over 2}.
\end{eqnarray}

From the above equation one can see that the average fidelity $F$ is
solely determined by the time evolution operator $\tilde{U}(t)$. In
a previous work \cite{Ref43}, Bowdrey et al. have also derived a
similar expression for average fidelity in the form of unitary (or
anti-unitary) operators. Moreover, Eq. (17) can be simplified under
certain special conditions. For example, if one chooses the
correlated spin chain Hamiltonian of the considered quantum system
as $\hat{H}=\sum_n[J_{n,n+1}(\sigma_n^x\sigma_{n+1}^x+
\sigma_n^y\sigma_{n+1}^y)+\Delta_{n,n+1}\sigma_n^z\sigma_{n+1}^z+
B_n\sigma_n^z]$, then it is direct to show that the last column of
$\hat{H}$ can be written as $\hat{H}_{:,d}=(0,0,\ldots,0,c)^T$ in
the standard basis, where $d$ is the dimension of $\hat{H}$,
$c=\sum_n(\Delta_{n,n+1}+B_n)$ is a nonzero number. Combination of
these results with the expression of the operator
$\tilde{\Lambda}=A\otimes I+I\otimes B'+\gamma\sum_{n}c_{n}\otimes
c_n^*$, one can show that the last column of $\tilde{\Lambda}$ has
the form $\tilde{\Lambda}_{:,d^2}=(0,0,\ldots,0)^T$, thus by
adopting the Taylor series expansion of the time evolution operator
$\tilde{U}(t)=\exp({\tilde{\Lambda}t})$, one can compute the last
column of $\tilde{U}(t)$ as $\tilde{U}_{:,d^2}=(0,0,\ldots,0,1)^T$
(note that this expression is valid for all subspaces, when
restricted to the subspace $\mathcal {H}_{0\oplus1}$, it holds true
even for the system Hamiltonian with inhomogeneous $x$ and $y$
components), this gives rise to a simplification of Eq. (17) as
\begin{eqnarray}
 F&=&{\tilde{U}_{N^2+N+n,N^2+N+m}+\tilde{U}_{Nn+n,Nm+m}\over 6}\nonumber\\&&
     +{\tilde{U}_{Nn-N+2n-1,Nm-N+2m-1}\over 6}+{1\over 2}.
\end{eqnarray}

Eqs. (17) and (18) are the main results we obtained in this paper,
which will be used in the latter discussion of quantum state
transfer when an explicit form of the spin chain Hamiltonian is
given. Particularly, for the mirror-symmetric Hamiltonian
$\hat{H}_m$ (by mirror symmetry, we mean that the interaction
Hamiltonian of the spin chain has symmetric coupling strengths about
the centre qubit or the centre link, i.e., the coupling strength
$J_{m,n}=J_{N-m+1,N-n+1}$) we have
$\tilde{U}_{N^2+N+n,N^2+N+m}=\tilde{U}_{Nn+n,Nm+m}^\dag$. In the
absence of decoherence environments (i.e., $\gamma=0$), this yields
$\tilde{U}_{N^2+N+n,N^2+N+m}+\tilde{U}_{Nn+n,Nm+m}\\=2{\rm
Re}\{\langle n|\exp(-i\hat{H}_m t)|m\rangle\}=2|\langle
n|\exp(-i\hat{H}_m t)|m\rangle|\cos\alpha$ and
$\tilde{U}_{Nn-N+2n-1,Nm-N+2m-1}=|\langle n|\exp(-i\hat{H}_m
t)|m\rangle|^2$, where $\alpha=\arg \{\langle n|\exp(-i\hat{H}_m
t)|m\rangle\}$ denotes the argument of the complex number $\langle
n|\exp(-i\hat{H}_m t)|m\rangle$. Clearly, this is just that of Eq.
(6) in Ref. \cite{Ref1}, which describes average fidelity in the
non-disturbed case.

\section{State transfer in decoherence spin channels}
\label{sec:3} Recently, intense research efforts has been devoted to
the interacting spin systems which were proposed as potential
candidates to simulate the relation between qubits in a quantum
computer \cite{Ref44}. This choice is due to the fact that such
systems can be easily manipulated (e.g., by tunneling potentials or
energy bias) and scaled up to large registers. In the present paper,
we investigate state transfer in this kind of system, with the
addition of the presence of the dissipative and dephasing
environments. We concentrated on the spin channel with fixed but
different coupling strengths between neighboring qubits, with the
Hamiltonian given by
\begin{equation}
 \hat{H}=\sum_{n=1}^{N-1}{J_{n,n+1}\over
 2}(\sigma_n^x\sigma_{n+1}^x+\sigma_n^y\sigma_{n+1}^y),
\end{equation}
where $\sigma_n^\alpha$ $(\alpha=x,y,z)$ are the usual Pauli
operators acting on the $n$th spin, $J_{n,n+1}=\lambda
\sqrt{n(N-n)}$ is the modulated exchange coupling strength, and
$\lambda$ is a scaling constant. This Hamiltonian is identical to
the representation of the Hamiltonian of a fictitious spin
$S=(N-1)/2$ particle: $\hat{H}_S=\lambda S_x$, where $S_x$ is its
angular momentum operator in the $x$-direction.

For this modulated spin chain, it has been shown that one can
achieve perfect state transfer between the input node $n$ and the
output node $N-n+1$ after a time $t_0=\pi/2\lambda$ and at intervals
of $\pi/\lambda$ thereafter in the absence of decoherence
environment, i.e., when $t=(2k-1)\pi/2\lambda$, $k=\{1,2,\ldots\}$,
the transfer fidelity reaches unity \cite{Ref3,Ref4}. When the
decoherence is present, however, this ideal communication channel
may be destroyed. To show that this is so, we consider effects of
dissipative and dephasing environments on quantum state transfer
from one end of the chain to another. The average fidelity can be
obtained directly from Eq. (18) by replacing $(m, n)$ with $(1, N)$.
Moreover, since the Hamiltonian $\hat{H}$ expressed in Eq. (19) is
mirror symmetric, we have
$\tilde{U}_{N^2+2N,N^2+N+1}=\tilde{U}_{N^2+N,N+1}^\dag$, thus the
average fidelity can be rewritten as
\begin{equation}
 F={|\tilde{U}_{N^2+2N,N^2+N+1}|\cos\alpha\over 3}
     +{\tilde{U}_{N^2+N-1,1}\over 6}+{1\over 2}.
\end{equation}

Physically, one can maximize the average fidelity $F$ by performing
a phase flip operation or applying an external magnetic field along
the $z$ axis such that $\alpha$ is a multiple of $2\pi$.
Furthermore, for the case of dissipative environment one can show
that $\tilde{U}_{N^2+N-1,1}=|\tilde{U}_{N^2+2N,N^2+N+1}|^2$, thus
the expression of the average fidelity $F$ is very similar to that
of Eq. (6) in Ref. \cite{Ref1}, which describes the average fidelity
in the non-disturbed case. However, for the case of dephasing
environment, one cannot obtain such a simple relation. Except this,
there exists another difference between the dissipative and the
dephasing environments which may be useful in the following
discussion. To demonstrate this explicitly, it is illuminating to
rewrite the ancillary operator $\tilde{\Lambda}=A\otimes I+I\otimes
B'+\gamma\sum_{n}c_{n}\otimes c_n^*$ as
$\tilde{\Lambda}=\tilde{\Lambda}(1)+\tilde{\Lambda}(2)$, where
$\tilde{\Lambda}(1)=-i(\hat{H}\otimes I-I\otimes\hat{H'})$ and
$\tilde{\Lambda}(2)=\gamma\sum_n\{c_n\otimes c_n^*-{1\over
2}[(c_n^\dag c_n)\otimes I+I\otimes(c_n^\dag c_n)']\}$, then for the
case of dissipative environment, one can show directly that
$\tilde{\Lambda}(1)$ and $\tilde{\Lambda}(2)$ commutes with each
other, i.e.,  $[\tilde{\Lambda}(1),\tilde{\Lambda}(2)]=0$, while for
the case of dephasing environment, one still cannot obtain this
commutation relation.

As a heuristic analysis, we first consider dissipative effects on
fidelity of quantum state transfer from one end of the chain to
another at time $t_0=(2k-1)\pi/2\lambda$. This choice of time $t_0$
is based on the fact that for weak system-environment coupling
strength $\gamma$, the difference between $t_0$ and the critical
time $t_c$ at which the transfer fidelity attains its maximum value
is minimal (e.g., for $N=2$ and $\gamma=0.1$, the deviation for
transfer fidelity of an excitation at time $t_0$ is about $0.2496\%
$ from that at time $t_c$), particularly, for large $N$ and small
$\gamma$, this difference can even be neglected (see the inset of
Fig. 1). For this kind of decoherence mode, from the commutation
relation $[\tilde{\Lambda}(1),\tilde{\Lambda}(2)]=0$ one can find
that the time evolution operator
$\tilde{U}(t)=\exp(\tilde{\Lambda}t)$ can be rewritten as
$\tilde{U}(t)=\exp[\tilde{\Lambda}(1)t]\exp[\tilde{\Lambda}(2)t]$.
For the first term, from the equality $(A\otimes L)(B\otimes
M)=AB\otimes LM$ and the Taylor series expansion of it one can
obtain $\exp[\tilde{\Lambda}(1)t]=[\exp(-it\hat{H})\otimes
I][I\otimes\exp(it\hat{H'})]$. When restricted to the zero and
single excitation subspace $\mathcal {H}_{0\oplus 1}$, the
eigenvalues and the corresponding eigenvectors of $\hat{H}$ can be
readily obtained as \cite{Ref45}
\begin{eqnarray}
 &\mathcal {E}_0=0,\quad \mathcal {E}_k=-(N-2k+1)\lambda,\nonumber\\
 &|\psi_0\rangle=|\textbf{0}\rangle,\quad
  |\psi_k\rangle=\sum_{n=1}^{N}c_{k,n}|n\rangle,
\end{eqnarray}
where $k=\{1,2,\ldots,N\}$, and the coefficient $c_{k,n}$ is given
by the following recursion relations
\begin{eqnarray}
 c_{1,1}&=&{1\over \sqrt{2^{(N-1)}}},\quad
 c_{k,1}=(-1)^{k+1}c_{1,k},\nonumber\\
 c_{k,n}&=&{\mathcal{E}_{k}c_{k,n-1}-
 \sqrt{(n-2)(N-n+2)}c_{k,n-2}\over\sqrt{(n-1)(N-n+1)}}\quad
 (n\geqslant2).\nonumber\\
\end{eqnarray}
Then $\exp[\tilde{\Lambda}(1)t_0]$ can be obtained analytically (for
the ease of presentation, we omitted its explicit expression since
this provides no new insights on the system), and the $(N^2+N-1)$-th
row of it is $[1,0,0,\ldots,0]$, while the $(N^2+2N)$-th row of it
is $[0,\ldots,0,i^{(N-1)(2k-1)},0,\ldots,0]$, with the only nonzero
element $i^{(N-1)(2k-1)}$ locating at the $(N^2+N+1)$-th column.

On the other hand, the eigenvalues and the eigenvectors of
$\tilde{\Lambda}(2)$ can also be obtained analytically, the results
are expressed as follows
\begin{eqnarray}
 &E=\gamma(0,c_1,c_1,\ldots,c_1,c_2),\nonumber\\
 &M=\left(\begin{array}{ccccc}
     &\quad g &\quad   &\quad        & \\
     &\quad   &\quad g &\quad        & \\
     &\quad   &\quad   &\quad \ddots & \\
     &\quad   &\quad   &\quad        &\quad g \\
   1 &\quad d &\quad d &\quad \cdots &\quad d
  \end{array}\right),
\end{eqnarray}
where the elements of the $(N+1)^2$-dimensional row vector $E$ are
eigenvalues of $\tilde{\Lambda}(2)$, $M$ is the $[(N+1)^2\times
(N+1)^2]$-dimensional matrix with eigenvectors of
$\tilde{\Lambda}(2)$ as its columns. Except the parameters $1$, $d$
and $g$, all the other elements in the matrix $M$ are zero. The
explicit form of $c_1$, $c_2$, $d$ and $g$ are given by
\begin{eqnarray}
 &c_1=-(\underbrace{1,1,\ldots,1}_N,0.5),\quad
 c_2=-(\underbrace{0.5,0.5,\ldots,0.5}_N),\nonumber\\
 &d=(-1/\sqrt{2},\underbrace{0,0,\ldots,0}_{N+1}), \quad
 g=diag(1/\sqrt{2},\underbrace{1,1,\ldots,1}_{N+1}).\nonumber\\
\end{eqnarray}
Thus the analytical form of $\exp[\tilde{\Lambda}(2)t_0]$ can also
be obtained. The first and the $(N^2+N+1)$-th column of it can be
written explicitly as $[e^{-\gamma t_0},0,\ldots,0,1-e^{-\gamma
t_0}]^T$ and $[0,\ldots,0,e^{-\gamma t_0/2},0,\ldots,0]^T$, with the
only nonzero element $e^{-\gamma t_0/2}$ locating at the
$(N^2+N+1)$-th row.

From the above analytical results, one can obtain the transfer
fidelity of an excitation and the average fidelity at time
$t_0=(2k-1)\pi/2\lambda$ as
\begin{eqnarray}
 f(t_0)&=&e^{-\gamma t_0}, \nonumber\\
 F(t_0)&=&{e^{-\gamma t_0/2}\cos\alpha\over 3}+{e^{-\gamma t_0}\over
 6}+{1\over 2},
\end{eqnarray}
where $\cos\alpha={\rm Re}[i^{(N-1)(2k-1)}]$. From the above
equation one can see that when the system subjects to dissipative
environment, the transfer fidelity $f(t_0)$ of an excitation is
independent of the spin chain length $N$. For fixed $t_0$, it will
exponentially decay with the increase of the decoherence rate
$\gamma$, while for fixed $\gamma$, it will exponentially decay with
the increase of $t_0$ (this phenomenon may be understood physically
from that fact that we have assumed the dissipation acts
independently and equally on each spin, and since there is a single
excitation, the survical probability of this excitation should be an
exponential function of the time $t_0$). The average fidelity
$F(t_0)$ is, however, dependent on the spin chain length $N$ due to
the existence of the phase factor $\cos\alpha$. As can be found from
Eq. (25), $F(t_0)$ displays three different dynamical patterns
corresponding to $N=4r+1$, $N=4r-1$ and $N=2r$ ($r=1,2,\ldots$). For
the latter two patterns of $N$, one needs to perform a phase flip
operation at the end spin belonging to the receiver \cite{Ref4} or
apply an external magnetic field to spins of the system \cite{Ref1}
to maximize the average fidelity (after these operations, the
average fidelity at time $t_0$ is also independent of the spin chain
length $N$ and will display behaviors completely the same as that
depicted by the red diamonds in the bottom panel of Fig. 1). If
there are no additional operations are performed, from Fig. 1 one
can observe that only when $N=4r+1$ (or equivalently, $N-1$ to be
divisible by 4) can the quantum spin channel may be superior to its
classical counterpart, for which the highest fidelity for
transmission of a quantum state is $2/3$ \cite{Ref46}.
\begin{figure}
\centering
\resizebox{0.45\textwidth}{!}{%
\includegraphics{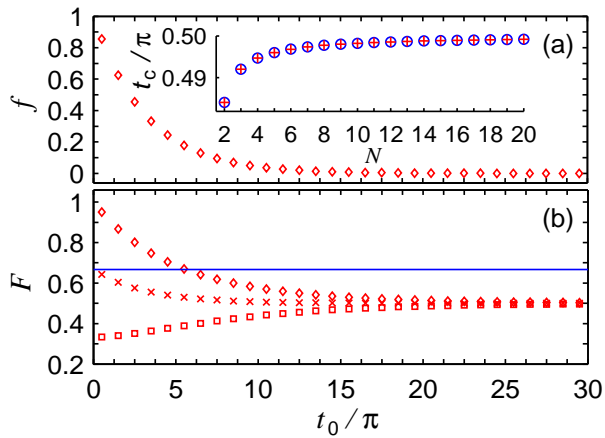}}
\caption{(Color online) Transfer fidelity of an excitation (a) and
the average fidelity (b) versus the rescaled time $t_0/\pi=(2k-1)/2$
($k=1,2,\ldots$) with the system subjects to dissipative
environment, where the decoherence rate and the scaling constant are
given by $\gamma=0.1$ and $\lambda=1$. In (b), the plots from top to
bottom correspond to $N=4r+1$, $N=2r$ and $N=4r-1$ ($r=1,2,\ldots$),
with the straight line at $F=2/3$ shows the highest fidelity for
classical transmission of a quantum state. The inset in (a) shows
$t_c/\pi$ versus $N$, $t_c$ is the critical time at which the
transfer fidelity of an excitation (blue circles) and the average
fidelity (red plus signs) attain their maxima.} \label{fig:1}
\end{figure}

For arbitrary evolution time $t$, after a tedious computation, we
obtain the state transfer fidelity of an excitation and the average
fidelity analytically as
\begin{eqnarray}
 f(t)&=&e^{-\gamma t}\sin^{2(N-1)}\lambda t, \nonumber\\
 F(t)&=&{e^{-\gamma t/2}\sin^{N-1}(\lambda t)\cos\alpha\over 3}+
 {e^{-\gamma t}\sin^{2(N-1)}\lambda t\over
 6}+{1\over 2}.\nonumber\\
\end{eqnarray}

From the above equation one can see that under the detrimental
influence of dissipative environment, both $f(t)$ and $F(t)$ behave
as suppressed oscillations as the time evolves. Moreover, the decay
of the average fidelity $F(t)$ displays three different patterns
mediated by a phase factor $\cos\alpha={\rm Re}[i^{(N-1)(2k-1)}]$.

From Eq. (26) one can also see that there exists a region in which
the dissipative spin channel is superior to its classical
counterpart. This region can be obtained analytically by solving the
nonlinear equation $F(t_c)>2/3$, which gives rise to
$\gamma<\ln(2\cos^2\alpha+2\cos\alpha\sqrt{1+\cos^2\alpha}+1)/t_c$.
If we choose $\cos\alpha=1$ and $t_c\simeq t_0=\pi/2$ (i.e.,
$\lambda=1$), then we have $\gamma<1.122$, and this result is
independent of the spin chain length $N$.
\begin{figure}
\centering
\resizebox{0.45\textwidth}{!}{%
\includegraphics{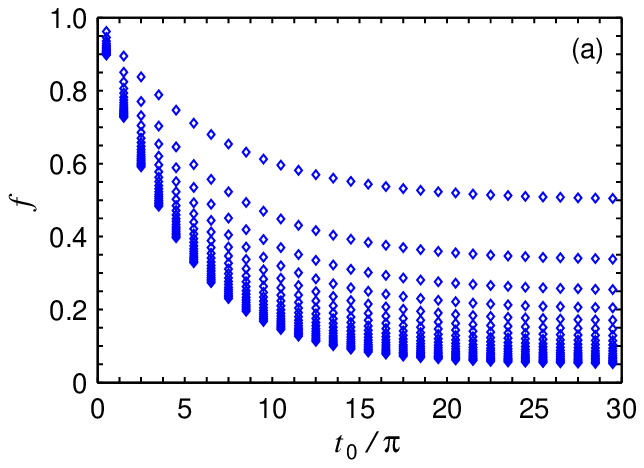}}
\resizebox{0.45\textwidth}{!}{%
\includegraphics{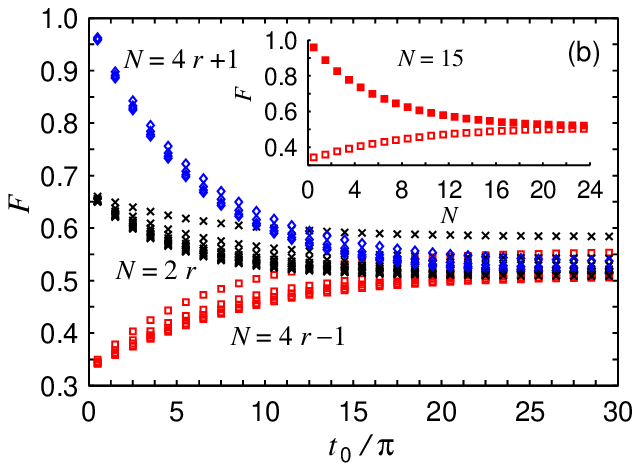}}
\caption{(Color online) Transfer fidelity of an excitation (a) and
the average fidelity (b) versus the rescaled time $t_0/\pi=(2k-1)/2$
($k=1,2,\ldots$) with the system subjects to dephasing environment.
Here the decoherence rate and the scaling constant are given by
$\gamma=0.1$ and $\lambda=1$. For (a), the plots from top to bottom
correspond to $N=2$ to $N=20$. For (b), the blue diamonds from top
to bottom correspond to $r=1$ to $r=4$, the black crosses from top
to bottom correspond to $r=1$ to $r=10$, and the red squares from
top to bottom correspond to $r=1$ to $r=5$. The inset in (b) is an
exemplified figure plotted for the comparison of the average
fidelity $F$ before (hollow squares) and after (solid squares) the
phase shift caused by the factor $\cos\alpha$ being corrected.}
\label{fig:2}
\end{figure}

When the system subjects to dephasing environment, however, one
cannot obtain an analytical result as that of the dissipative
environment because $[\tilde{\Lambda}(1), \tilde{\Lambda}(2)]\neq
0$, so we resort to numerical techniques. The typical results are
shown in Fig. 2, from which one can see that the transfer fidelity
of an excitation $f(t_0)$ also decays with the increase of $t_0$,
however, when $t_0\rightarrow\infty$, $f(t_0)$ arrives at a nonzero
value $f(t_0\rightarrow\infty)=1/N$, and this is different from that
of the dissipative environment, in which $f(t_0)$ decays to zero
asymptotically. There is another difference exists, namely, for any
fixed decoherence rate $\gamma$, $f(t_0)$ decreases with the
increase of $N$ for the case of dephasing environment, which is
disadvantageous for long distance quantum communication. The average
fidelity $F(t_0)$ is also dependent on the spin chain length, and it
displays three different patterns for different kinds of $N$. As can
be seen from Fig. 2b, only for the case of $N=4r+1$ ($r=1,2,\ldots$)
can the quantum spin channel may be superior to its classical
counterpart if no other performance have been made. When
$t_0\rightarrow\infty$, the average fidelity also arrives at a
constant value which is also dependent on $N$, and can be computed
analytically as $F(t_0\rightarrow\infty)=1/6N+1/2$.
\begin{figure}
\centering
\resizebox{0.45\textwidth}{!}{%
\includegraphics{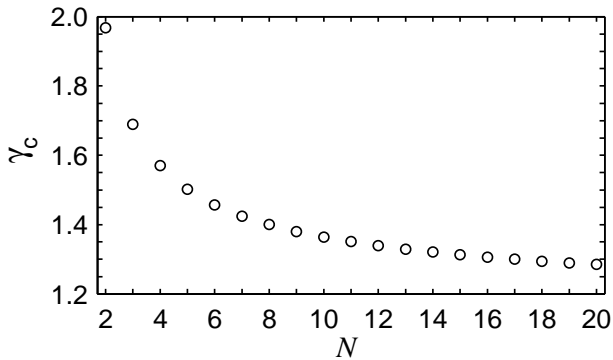}}
\caption{Critical decoherence rate $\gamma_c$ versus $N$ with the
system subjects to dephasing environment, where the scaling constant
$\lambda$ has been chosen to be 1, and the phase shift for $N=4r-1$
and $N=2r$ has been assumed to be eliminated.} \label{fig:3}
\end{figure}

For fixed spin chain length $N$ and critical time $t_c$, there also
exists a critical decoherence rate $\gamma_c$ before which the
dephasing channel is superior to any classical communication
channel, for which can only ensure transmission of a quantum state
with the highest fidelity $2/3$ \cite{Ref46}. The typical plots of
$\gamma_c$ versus $N$ with scaling constant $\lambda=1$ is shown in
Fig. 3, where we have assumed that the phase shift caused by the
factor $\cos\alpha$ is eliminated by choosing $\cos\alpha=1$ for the
cases of $N=4r-1$ and $N=2r$. From this figure one can see that
$\gamma_c$ decreases with the increase of $N$, this puts another
constraint on this spin channel for long distance quantum
communication because for large $N$, one needs to reduce the mutual
interaction between the system of interest and its surrounding
environment in order to obtain a reliable transfer fidelity, and in
general, this is not an easy task in real experiments.
\begin{figure}
\centering
\resizebox{0.45\textwidth}{!}{%
\includegraphics{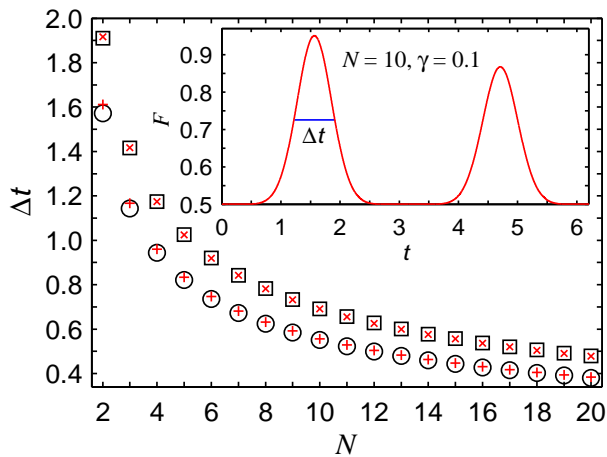}}
\caption{(Color online) FWHM $\Delta t$ of the first oscillation
versus $N$ with $\gamma=0.1$ and $\lambda=1$. The black squares and
circles show $\Delta t$ corresponds to the case of the transfer
fidelity of an excitation and the average fidelity with the system
subjects to dissipative environment, while the red crosses and plus
signs show $\Delta t$ corresponds to those with the system subjects
to dephasing environment. Moreover, the average fidelity was plotted
with the phase shift being corrected, and the inset is a schematic
picture (with the system subjects to dissipative environment)
annotating the meaning of the FWHM.} \label{fig:4}
\end{figure}

From the aforementioned analysis one can see that increasing the
decoherence rate $\gamma$ or the spin chain length $N$ always
introduces stringent constraints on the spin channel for long
distance quantum communication. In the following, we discuss another
serious constraint imposed by the dissipative and dephasing
environments on the spin channel. For this purpose, we examine full
width at half maximum (FWHM) of the first oscillation describing the
dynamics of the fidelity of an excitation as well as the average
fidelity and we denote it as $\Delta t=|t_2-t_1|$, with
$f(t_1)=f(t_2)=f_{max}/2$ and $F(t_1)=F(t_2)=F_{max}/2$. For the
case of dissipative environment, from Eq. (26) one can obtain the
following two relations
\begin{eqnarray}
 e^{-\gamma t}\sin^{2(N-1)}\lambda t&=&f_{max}/2,\nonumber\\
 e^{-\gamma t/2}\sin^{N-1}\lambda
 t&=&\sqrt{3F_{max}-3+\cos^2\alpha}-\cos\alpha,\quad
\end{eqnarray}
where for transfer fidelity of an excitation, $(t_1, t_2)$ can be
obtained by solving the nonlinear equation expressed in the first
line of Eq. (27), and for the average fidelity, $(t_1, t_2)$ can be
obtained by solving the second equation expressed in Eq. (27). The
maximum value of $f_{max}$ and $F_{max}$ can be obtained directly
from Eq. (26).

For the case of dephasing environment, however, one cannot obtain a
result similar to that of Eq. (27), thus we resort to numerical
methods. The exemplified plots are shown in Fig. 4, where for the
case of average fidelity, we have chosen $\cos\alpha=1$. From this
figure one can see that the curves exhibit nearly the same behaviors
for both of the decoherence scenarios. With the increase of the spin
channel length $N$, the FWHM $\Delta t$ decreases one by one, which
implies that the peak of the wave becomes steeper and narrow with
the increase of $N$. This puts a stringent limitation on the time
when the measurement must be performed by the receiver Bob, because
for large $N$, when the transmitted state arrives at the destination
qubits, Bob should make measurements on the qubits at the instant
time of $t_c$ as quickly as possible, otherwise, even a minor
deviation of measurement time from $t_c$ would lead to a great
inaccuracy to the transmitted state.

We now turn to investigate quantum state transfer in another spin
channel proposed by Shi et al. \cite{Ref6}, the Hamiltonian of the
system has the similar form as that expressed in Eq. (19), with
however, the coupling strength between neighboring spins given by
\begin{equation}
 J_{n,n+1}=\left\{
    \begin{aligned}
         &\lambda\sqrt{n(N-n)}\quad (n\in {\rm even}),\\
         &\lambda\sqrt{(n+2k)(N-n+2k)}\quad (n\in {\rm odd}).
    \end{aligned} \right.
\end{equation}
Here $k=\{0,1,2,\ldots\}$, and $\lambda$ is still a scaling
constant. This Hamiltonian is the same as that of Eq. (19) when
$k=0$. For other cases of $k$ and $N\in {\rm even}$, it can be used
to achieve perfect state transfer from the first node to the last
node at time $t_0=\pi/2\lambda$ and periodically returns there at
regular time intervals of $\pi/\lambda$ in the absence of
decoherence environment (for the special case of $N=2$, perfect
state transfer can also be achieved at time $t=\pi/(4k+2)\lambda+\pi
r/(2k+1)\lambda$, $r=\{0,1,2,\ldots\}$). Here we discuss how the
dissipative and dephasing environments affect quantum state transfer
in this spin channel. Since the case of $k=0$ has been discussed in
the above section, here we concentrated only on dynamics of state
transfer fidelity at time $t_0=(2k-1)\pi/2\lambda$ for the case of
$k\geqslant 1$. The eigenvalues and the eigenvectors of the
corresponding Hamiltonian have been obtained in Ref. \cite{Ref6},
and this enables an analytical analysis for the case of dissipative
environment, while for dephasing environment, we also resort to
numerical analysis. Our results revealed that for both of the
decoherence modes, the transfer fidelity of an excitation and the
average fidelity at time $t_0$ for $k\geqslant 1$ show completely
the same behaviors as that of $k=0$ and $N\in {\rm even}$, thus the
region in which this spin channel is superior to any classical
communication channel is also the same as that of $k=0$ and $N\in
{\rm even}$. However, for any fixed $N\in {\rm even}$, the FWHM
$\Delta t$ becomes narrower and narrower with the increase of $k$,
thus by the same argument as made in the previous section, we
suggest that in practice, one can choose $k=0$ in order to minimize
the deviations from the transmitted state introduced by the
measurement process.

Moreover, for the special case of $N=2$ at arbitrary evolution time
$t$, the transfer fidelity of an excitation and the average fidelity
with the system subjects to dissipative environment can be obtained
analytically as $f(t)=e^{-\gamma t}\sin^2[(2k+1)\lambda t]$ and
$F(t)=e^{-\gamma t}\sin^2[(2k+1)\lambda t]/6+1/2$. For the case of
the system subjects to dephasing environment, again due to the fact
that $[\tilde{\Lambda}(1),\tilde{\Lambda}(2)]\neq 0$, one cannot
obtain an analytical result of the state transfer fidelity.

\section{Entanglement distribution and creation in decoherence spin channels}
\label{sec:4} From the above discussion one can see that both the
dissipative and dephasing environments have severely detrimental
effects on fidelity of quantum state transfer. How these decoherence
modes affect other dynamical processes of a quantum system. As an
answer to this question, in this section we investigate decoherence
effects on entanglement distribution and creation. We first see
entanglement distribution between two distant parties through the
dissipative and dephasing spin channels. To assess the extent to
which these decoherence environments affect entanglement
distribution, we assume the entangled state
$|\psi\rangle=(|01\rangle+|10\rangle)/\sqrt{2}$ is initially
prepared between a non-interacting qubit NI and the first qubit A of
the chain, then after some time $t$, the entanglement may be
established between NI and the $N$th qubit B. The overall
Hamiltonian of the system can be written as
$\hat{H}_{over}=I_2\otimes \hat{H}$ ($I_2$ denotes the $2\times 2$
identity matrix). To assess the amount of the pairwise entanglement
at different instants of time, we adopt the concurrence, a function
introduced by Wootters \cite{Ref47}, equals to 1 for maximally
entangled states and zero for separable states, defined as
$C=\max\{0,\lambda_1-\lambda_2-\lambda_3-\lambda_4\}$, where
$\lambda_i$ ($i=1,2,3,4$) are the square roots of the eigenvalues of
the matrix $R=\rho_{mn}(\sigma_m^y\otimes
\sigma_n^y)\rho_{mn}^*(\sigma_m^y\otimes \sigma_n^y)$ ($\rho_{mn}$
is the two-qubit reduced density matrix). Then after a similar
numerical analysis as performed in Section 3, we obtained
entanglement dynamics of the system. The results revealed two
remarkable features (see Fig. 5). First, for both of the decoherence
modes, the pairwise entanglement between the non-interacting qubit
NI and the last qubit B measured by the concurrence $C_{\rm
NI,B}(t)$ display completely the same behaviors, i.e., they both
behave as suppressed fluctuations as the time $t$ evolves, and the
entanglement may experience sudden death \cite{Ref48} during certain
intervals of the evolution time (see the inset in the top panel of
Fig. 5). Second, for any fixed decoherence rate $\gamma$, the
maximum value of $C_{\rm NI,B}(t)$ decreases smoothly with the
increase of the spin chain length $N$, and the critical time $t_c$
at which the concurrence $C_{\rm NI,B}(t)$ attains its maximum value
still does not locate at $t_0=\pi/2\lambda$ (at this point, the
concurrence can be obtained analytically as $C_{\rm
NI,B}(t_0)=e^{-\gamma t_0}$, which is independent of $N$). $t_c$
increases as $N$ increases, and when $N\rightarrow\infty$, $t_c$
approaches its asymptotic value $\pi/2\lambda$, which is the same as
that of $t_0$. This implies that even for very large $N$, one can
still obtain substantial concurrence $C_{\rm
NI,B}(t_c)=e^{-\gamma\pi/2\lambda}$ (it is not difficult to
understand this phenomenon, because the scaling for large $N$ also
involves increasing the coupling strengths between the neighboring
spins for we have chosen $J_{n,n+1}=\lambda\sqrt{n(N-n)}$).
\begin{figure}
\centering
\resizebox{0.45\textwidth}{!}{%
\includegraphics{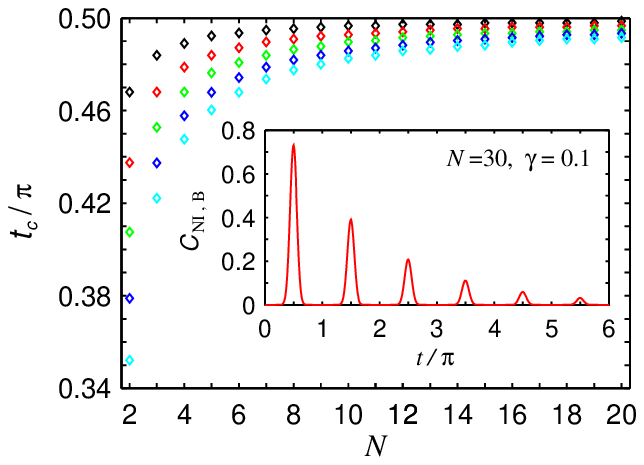}}
\resizebox{0.45\textwidth}{!}{%
\includegraphics{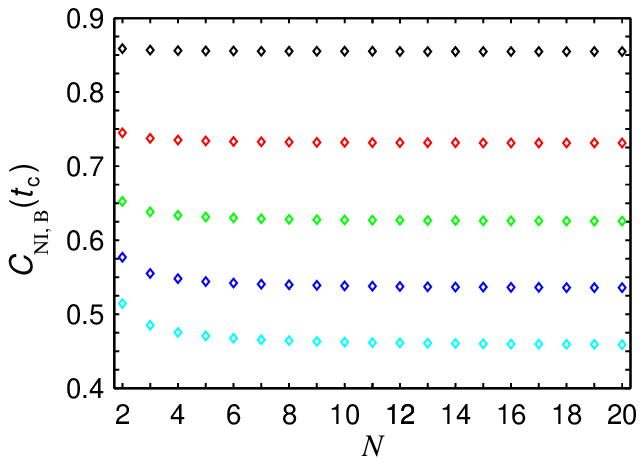}}
\caption{(Color online) Rescaled critical time $t_c/\pi$ and the
maximum concurrence $C_{\rm NI,B}(t_c)$ versus $N$ with the system
subjects to dissipative or dephasing environment. For every plot,
the diamonds from top to bottom correspond to $\gamma=0.1, 0.2, 0.3,
0.4, 0.5$. The inset in the top panel shows dynamics of the
concurrence $C_{\rm NI,B}$ versus $t/\pi$. The other parameter
values are given by $\gamma=0.1$ and $\lambda=1$.} \label{fig:5}
\end{figure}

Next we turn to investigate how the dissipative and dephasing
environments affect the creation of entanglement in spin networks.
For this purpose, we consider the multiarm structure
$M(N_1,N_2,N_A)$ of the XX spin chain Hamiltonian (19) with the
addition of the exchange couplings between the hub site and its
nearest-neighbor output sites satisfy the branching rule
\cite{Ref10,Ref49}. Here $N_1$ and $N_2$ denote the number of sites
in the input and output arms, respectively, and $N_A$ denotes the
number of output arms (see the inset in the top panel of Fig. 6). It
has been shown that in the absence of decoherence environment, this
structure can be employed to create multi-qubit entangled \textit{W}
state \cite{Ref10,Ref50} at the ends of the outgoing arms. When the
system is subjected to the decoherence environment, however, the
case may be different.

When the quantum system subjects to dissipative environment, our
numerical results revealed that the concurrence $C$ (here $C$
measures amount of the created pairwise entanglement between
arbitrary two end nodes of the output arms) is determined by two
numbers $N=N_1+N_2+1$ and $N_A$. For any fixed $N_A$ and decoherence
rate $\gamma$, the maximum value of the concurrence decreases with
the increase of $N$. This is different from that of the ideal case
(i.e., $\gamma=0$), for which the concurrence has no relation with
$N$ \cite{Ref10,Ref49}. Moreover, similar to the case of
entanglement distribution in a spin chain, the critical time $t_c$
at which the concurrence attains its maximum value is also slightly
different from $t_0=\pi/2\lambda$ for small values of $N$ (see the
top panel of Fig. 6), while $t_c$ is nearly independent of $N_A$. In
the limit of $N\rightarrow\infty$, however, $t_c\rightarrow
t_0=\pi/2\lambda$, at which the concurrence can be obtained
explicitly as $C(t_c)=2e^{-(\gamma\pi/2\lambda)}/N_A$.
\begin{figure}
\centering
\resizebox{0.45\textwidth}{!}{%
\includegraphics{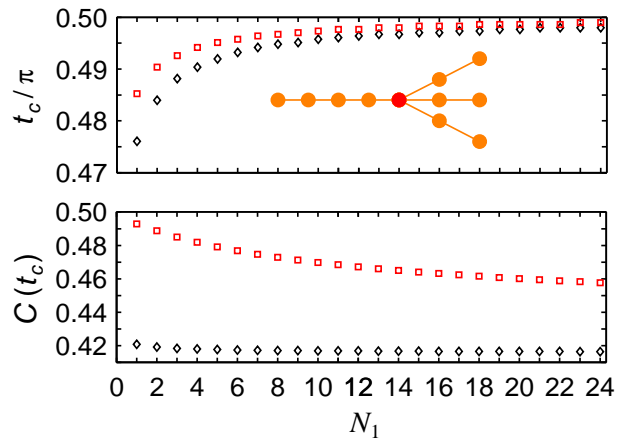}}
\caption{(Color online) Rescaled critical time $t_c/\pi$ and the
maximum concurrence $C(t_c)$ versus $N_1$ with the system subjects
to dissipative (black diamonds) and dephasing (red squares)
environments. The other parameters for the plots are $N_2=1$,
$N_A=3$, $\gamma=0.3$ and $\lambda=1$. The inset in the top panel is
an exemplified multiarm structure of the spin network with the solid
red circle denotes the hub, and the number of sites in the input and
output arms are given by $N_1=4$, $N_2=2$, while the number of
output arms is $N_A=3$.} \label{fig:6}
\end{figure}

When the system subjects to dephasing environment, however, our
numerical results revealed that the concurrence $C(t)$ is dependent
on $N_1$, $N_2$ and $N_A$, which is different from that of the
dissipative environment. For fixed $N=N_1+N_2+1$, the maximum value
of $C(t)$ decreases with the increase of $N_2$, thus in order to
minimize the detrimental effects of dephasing environment on
entanglement creation, one can design the spin structure with the
number of sites in each output arm as $N_2=1$. Moreover, as can be
seen from Fig. 6, when the system subjects to dephasing environment,
the critical time $t_c$ at which the concurrence attains its maximum
approaches the asymptotic value $\pi/2\lambda$ more quickly than
that for the dissipative environment, and the detrimental effects of
the dissipative environment seems to be more severe than that of the
dephasing environment under the condition of the same decoherence
rate $\gamma$.

\section{ Summary and discussion}
\label{sec:5} In the present paper, by exact diagonalization of a
generalized superoperator for solving the master equation, we
investigated quantum state transfer, as well as entanglement
distribution and creation through the dissipative and dephasing spin
channels, aimed at revealing the extent to which these decoherence
environments affect various QIP tasks. We focused on several
interaction-modulated spin networks which may serve as perfect spin
channels in the absence of decoherence, and gave an analytical
analysis for the case of dissipative environment, and numerical
analysis for the case of dephasing environment. The results revealed
three general conclusions. The first is that these two decoherence
environments always lead to suppression of the state transfer
fidelity, as well as the amount of the distributed and created
pairwise entanglement, particularly, this suppression increases with
the increase of the decoherence rate $\gamma$. The second is that
the detrimental effects of the dissipative environment are more
severe than that of the dephasing environment under the condition of
the same decoherence rate $\gamma$. The third is that the critical
time $t_c$ at which the transfer fidelity and the concurrence attain
their maxima increases with the increase of $N$, and when
$N\rightarrow\infty$, it approaches the asymptotic value
$t_0=\pi/2\lambda$.

For the case of state transfer, the results also revealed that for
fixed $\gamma$, the transfer fidelity of an excitation $f(t_0)$
decreases with the increase of $t_0$, while the average fidelity
$F(t_0)$ displays three different patterns corresponding to the
chain length $N=4r+1$, $N=4r-1$ and $N=2r$, and only when $N=4r+1$
can the spin channel superior to all of its classical counterpart if
there are no additional operations being performed. Besides this,
when the system subjects to dissipative environment, $f(t_0)$ is
independent of $N$, while for each pattern $N$, $F(t_0)$ is
independent of $r$. When the system subjects to dephasing
environment, however, both $f(t_0)$ and $F(t_0)$ decrease with the
increase of $N$ and $r$. Moreover, the critical decoherence rate
$\gamma_c$ before which the decoherence spin channel is superior to
its classical counterpart is independent of $N$ for dissipative
environment, and it decreases with the increase of $N$ for dephasing
environment.

For entanglement distribution and creation, our results revealed
that the distributed entanglement shows completely the same
dynamical behaviors when the system subjects to dissipative or
dephasing environment, and for fixed $\gamma$, the maximum value of
the concurrence $C_{\rm NI,B}$ decreases as $N$ increases, and when
$N\rightarrow\infty$, $C_{\rm NI,B}(t_c)=e^{-\gamma\pi/2\lambda}$.
The created entanglement also decreases as $N$ increases, and for
the case of dissipative environment, the maximum concurrence $C$
approaches its asymptotic value
$C(t_c)=2e^{-(\gamma\pi/2\lambda)}/N_A$ in the limit of
$N\rightarrow\infty$, while for the case of dephasing environment,
its asymptotic value can only be obtained numerically.

From the above analysis it can be seen that both the dissipative and
dephasing environments have severe detrimental effects on transport
capacity of a spin channel, thus a question naturally arises at this
stage is how to stabilize a quantum system against this unwanted
phenomenon, or in other words, how to minimize, delay or even
eliminate the detrimental effects of decoherence. In a recent work
\cite{Ref51}, we have shown that for some specific environments, it
is possible to obtain a near perfect state transfer by adjusting the
coupling strengths between neighboring qubits of a spin network,
however, for decoherence modes considered here, this procedure
cannot be applied, thus one must resort to other techniques in order
to get a high efficient communication. Physically, one can increase
the transfer fidelity to some degree by increasing the coupling
strength between neighboring qubits of the spin network, or
equivalently, by increasing the scaling constant $\lambda$ (this
statement can be proved true by the explicit form of the operator
$\tilde{\Lambda}$, from which one can obtain that increasing
$\lambda$ will decrease the period of the suppressed oscillation and
thus enlarges the maximum value of the state transfer fidelity).
Particularly, in the limit of $\lambda\rightarrow\infty$, the
transfer fidelity will approaches unity. However, the realization of
$\lambda$ large enough to obtain high efficient communication might
be a difficult experimental task.

Finally, we would like to emphasize that although there are works
showing that efficiency of the quantum state transfer can sometimes
be enhanced to some extent by decoherence \cite{Ref52,Ref53,Ref54},
finding ways to minimize or even eliminate the detrimental effects
of decoherence is still the prerequisites and challenging task in
the practical realization of quantum computer.
\\ \\ \\
This work was supported by the Natural Science Foundation of Shaanxi
Province under Grant Nos. 2009JQ8006 and 2009GM1007, the Specialized
Research Program of Education Bureau of Shaanxi Province under Grant
No. 08JK434, and the Youth Foundation of Xi'an University of Posts
and Telecommunications under Grant No. ZL2008-11.

%
\newcommand{\PRL}{Phys. Rev. Lett. }
\newcommand{\PRA}{Phys. Rev. A }
\newcommand{\NJP}{New J. Phys. }
\newcommand{\PLA}{Phys. Lett. A }
\newcommand{\JPB}{J. Phys. B }
\newcommand{\EPJD}{Eur. Phys. J. D }
%

\end{document}